# Large magnetoelectric effect in the $Al_{1-x}Ga_xFeO_3$ family of multiferroic Oxides

Ajmala Shireen, Rana Saha, A. Sundaresan [*] and C. N. R. Rao [*]

[a] New Chemistry Unit, Chemistry and Physics of Materials Unit and International Centre for Materials Science, Jawaharlal Nehru Centre forAdvanced Scientific Research, Bangalore, 560064, India.

E-mail: cnrrao@jncasr.ac.in ; sundaresan@jncasr.ac.in Fax: +91-80-22082766; Tel: +91-80-22082761

**Abstract:**

$AlFeO_3$, $GaFeO_3$ and $Al_{0.5}Ga_{0.5}FeO_3$, all crystallizing in a non-centrosymmetric space group, are multiferroic and also exhibit magnetoelectric properties, with $Al_{0.5}Ga_{0.5}FeO_3$ showing unusually large magnetocapacitance at 300 K.

Multiferroic oxides are not common since it is difficult to satisfy the criterion required for magnetism and ferroelectricity in the same material.[1] However, there are a few materials which show multiferroic properties because of mechanisms different from those in conventional ferroelectrics such as $BaTiO_3$ . All multiferroic materials are not necessarily magnetoelectric because the latter property arises from the interaction between the magnetic and the electric order parameters.[1-3] It has been recently reported that $AlFeO_3$ , which is derived from $Fe_2O_3$ by the substitution of one $Fe^{3+}$ by $Al^{3+}$ , shows ferrimagnetic as well as ferroelectric properties.[4-6] $AlFeO_3$ has an orthorhombic structure with the chiral, non-centrosymmetric $Pna2_1$ as the space group . $GaFeO_3$ has a structure similar to that of $AlFeO_3$ with a non centrosymmetric space group.[7] $GaFeO_3$ is also ferrimagnetic and ferroelectric.[6-10] In both $AlFeO_3$ and $GaFeO_3$, there are four different cation sites Fe1, Fe2, A1 and A2 (A=Al, Ga), of which Fe1, Fe2 and A2 have an octahedral oxygen environment. The cation in the A1 site is in the tetrahedral oxygen environment. The octahedral environment of Fe1, Fe2 and A2 are slightly distorted, while the $A1O_4$ tetrahedron is almost regular .The cause for this local deformation of the lattice is attributed to the  difference in the octahedral radii of  $Fe^{3+}$ and $A^{3+}$ ions and the disorder in the occupation of cation sites , especially in the octahedral sites. Low temperature magnetic ordering in these compounds is due to the cation-oxygen-cation superexchange antiferromagnetic interaction.  Magnetoelectric properties of multiferroic $AlFeO_3$ and $GaFeO_3$ are not established.In this communication, we present the multiferroic and magneto - electric properties of AlFeO$_3$ and GaFeO$_3$.[‡]More importantly, we have investigated the properties of Al$_{0.5}$Ga$_{0.5}$FeO$_3$ since the nature of the A-cations as well as cation disorder would have a significant effect on dielectric and relatedproperties.[‡] We find that Al$_{0.5}$Ga$_{0.5}$FeO$_3$ is not only multiferroic, but also shows remarkably large magnetocapacitance at 300 K,excelling the performance of AlFeO$_3$ and GaFeO$_3$.

AlFeO$_3$ is ferrimagnetic with a T$_N$ of 250 K, and two samples of this oxide prepared by different methods confirm this feature (see Fig. 1a for typical magnetisation data). GaFeO$_3$ exhibits a ferrimagnetic T$_N$ of 210 K. Al$_{0.5}$Ga$_{0.5}$FeO$_3$ exhibits a magnetic behaviour similar to AlFeO$_3$ and GaFeO$_3$ with a T$_N$ of 220 K, (Fig. 1b).We see divergence between the field-cooled (FC) and zero-field-cooled (ZFC) magnetization data just as in AlFeO$_3$ or GaFeO$_3$.We observe magnetic hysteresis at low temperatures (see inset of Fig. 1b).The value of saturation magnetization, remanant magnetization , and coercive field in the case of Al$_{0.5}$Ga$_{0.5}$FeO$_3$ are 10.0 emu/g , 19.1 emu/g and 10.3 kOe respectively, not very different from the values found in AlFeO$_3$ and GaFeO$_3$.

We show the temperature variation of dielectric properties of GaFeO$_3$ and Al$_{0.5}$Ga$_{0.5}$FeO$_3$ in Figures 2(a) and (b) respectively. We see dielectric dispersion below T$_N$ and a significant increase in the dielectric constant above T$_N$, a behaviour similar to that found in AlFeO$_3$ .The dielectric behaviour shown in Fig. 2 is not unlike that of relaxor ferroelectrics. Al$_{0.5}$Ga$_{0.5}$FeO$_3$ shows ferroelectric hysteresis at relatively low temperatures (< 300 K) just as AlFeO$_3$ and GaFeO$_3$, with the loops indicating the leaky nature of these materials. The maximum (saturation) polarization (P$_m$) and remanant polarization (P$_R$) are much higher in Al$_{0.5}$Ga$_{0.5}$FeO$_3$ than in GaFeO$_3$ or AlFeO$_3$ the values being P$_s$ = 1.1 µC/ cm$^2$ , P$_R$ = 0.5 µC/cm$^2$ and coercive field (E$_c$) =3.8 kV/cm at 200 K for an applied voltage of 500 V compared to P$_m$ = 0.05(0.07) µC/cm$^2$ , P$_R$ = 0.02(0.04) µC/cm$^2$ and E$_c$ = 1.22(6.4) kV/cm found in AlFeO$_3$ (GaFeO$_3$). Clearly, Al$_{0.5}$Ga$_{0.5}$ FeO$_3$ is a multiferroic with better ferroelectric properties.

A marked increase in the dielectric constant around T$_N$ signifies interaction between electric and magnetic order parameters. Direct evidence for this is provided by the effect of magnetic fields on the dielectric properties. Application of a magnetic field has a marked effect on the dielectric properties of GaFeO$_3$ and Al$_{0.5}$Ga$_{0.5}$FeO$_3$, but the effect is considerably smaller in AlFeO$_3$ . We show the effect of a magnetic field of 2 T on the properties of GaFeO$_3$ and Al$_{0.5}$Ga$_{0.5}$FeO$_3$ in Figure 2. Magnetocapacitance in AlFeO$_3$ was small (~2%) and its variation with frequency is negligible. The effect of the magnetic field is however large in the case of GaFeO$_3$ and Al$_{0.5}$Ga$_{0.5}$FeO$_3$. Both these oxides shows large magnetocapacitance and a significant variation with frequency as shown in Fig. 3(a). Maximum magnetocapacitance is observed in these two oxides around 40 kHz and at this frequency, Al$_{0.5}$Ga$_{0.5}$FeO$_3$ shows 60% magnetocapacitance at 2 T and 35% at 1 T. Such a large large magnetocapacitance is indeed noteworthy. In Fig. 3(b), we show the variation of % magnetocapacitance with magnetic field for all three oxides. We clearly see negligible effect of the magnetic field in the case of AlFeO$_3$, but a marked variation in GaFeO$_3$ and Al$_{0.5}$Ga$_{0.5}$FeO$_3$, the %magnetocapacitance increasing with the magnetic field.

In conclusion, it is significant that sample iron oxides of the type AFeO$_3$ (A= Al, Ga) are both multiferroic and magnetoelectric. The large magnetoelectric effect exhibited by GaFeO$_3$ and Al$_{0.5}$Ga$_{0.5}$FeO$_3$, specially the latter demonstrates the importance of the A-site cations. Displacements of the A cations and the disorder associated with their occupancies markedly effect the dielectric and magnetoelectric properties in a significant manner.[6]

**Notes and References:**

1. N. A. Hill, *J. Phys. Chem. B*, 2000, **104**, 6694.

2. R. Ramesh, N. A. Spaldin, *Nature Mater.*, 2007, **6**, 21.

3. C. N. R. Rao and C. R. SerraoJ, *J. Mater. Chem.*, 2007, **17**, 4931.

4. F. Bouree, J. L Baudour, E . Elbadraoui, J. Musso, C . Laurent and A.Rousse *Acta. Cryst.*, 1996, **B 52**, 217 .

5. L. F. Cotica ,S.N. De Medeiros , I . A. Santos ,A. Paesano JR. , E. J .Kinast, J. B. M. Da Cunha, M. Venet, D. Garcia and J. A. Eiras, *Ferroelectrics*, 2006, **338**, 241.

   L. F. Cotica, I . A. Santos, M. Venet, D.Garcia, J. A. Eiras and A.A.Coelho, *Solid State Commun.*, 2008, **147**, 123.

6. E. Kan, H Xiang, C. Lee, F Wu, J Yang and M-H. Whangbo *Angew.Chem. Int. Ed.*, 2010, **49**, 1603.

7. S. C. Abrahams, J. M. Reddy, and J. L. Bernstein, *J. Chem. Phys.*, 1965, **42**, 3957.

8. K. U. Kang, S. B. Kim, S.Y. An, S-W. Cheong, C. S. Kim, *J. Magn. Magn. Mater.*, 2006, **304**, c769.

9. K. Sharma, V. R. Reddy, D. Kothari, A. Gupta, A. Banerjee and V.G. Sathe, *J. Phys. Condens. Matter*, 2010, **22**, 146005.

10. V. B. Naik and R. Mahendiran, *J. Appl. Phys.*, 2009, **106**, 123910.

‡ AlFeO$_3$ and GaFeO$_3$ were prepared by co-precipitation method and sintered at 1350° C. One sample of AlFeO$_3$ was also prepared by solid state method at 1400° C. Al$_{0.5}$Ga$_{0.5}$FeO$_3$ was prepared by heating the appropriate mixture of Al$_2$O$_3$, Ga$_2$O$_3$ and Fe$_2$O$_3$ at 1400°C for 5 hrs. Al$_{0.5}$Ga$_{0.5}$FeO$_3$ has the similar structure as AlFeO$_3$ and GaFeO$_3$.. ($a = 5.00, b = 8.66, c = 9.353$)

**Figure Captions:**

Figure 1: Temperature dependent magnetization of (a) $AlFeO_3$ and (b) $Al_{0.5}Ga_{0.5}FeO_3$ under field cooled (FC) and zero field cooled (ZFC) conditions. Magnetic hysteresis at 5 K is shown in the inset.

Figure 2: Temperature variation of dielectric properties of (a) $GaFeO_3$ and (b) $Al_{0.5}Ga_{0.5}FeO_3$. Effect of 2 T is also shown for both (a) and (b).

Figure 3: (a) Variation of % magnetocapacitance with frequency for $GaFeO_3$ and $Al_{0.5}Ga_{0.5}FeO_3$ at a magnetic field of 2 T. In the inset variation of % magnetocapacitance with frequency at 1 T is shown. (b) Variation of % magnetocapacitance with magnetic field for $AlFeO_3$ (1 kHz, 250 K), $GaFeO_3$ (40 kHz, 300 K) and $Al_{0.5}Ga_{0.5}FeO_3$ (40 kHz, 300 K). Data for two different preparations of $AlFeO_3$ are shown.

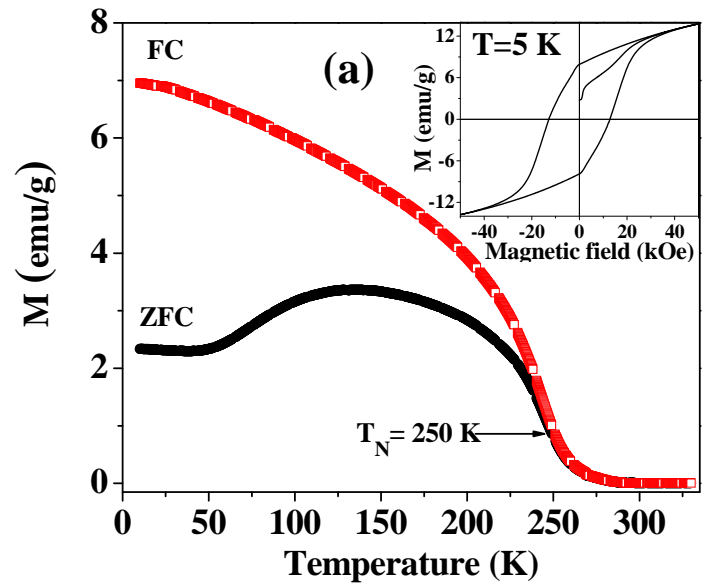
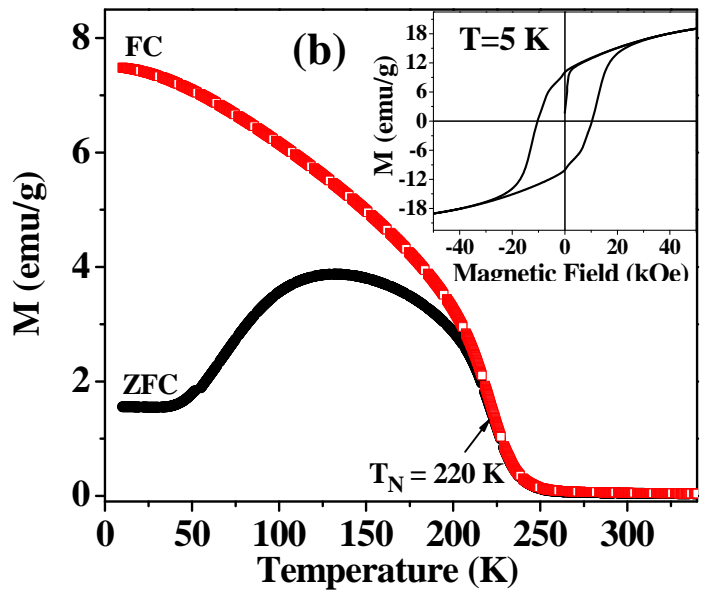

**Figure 1**

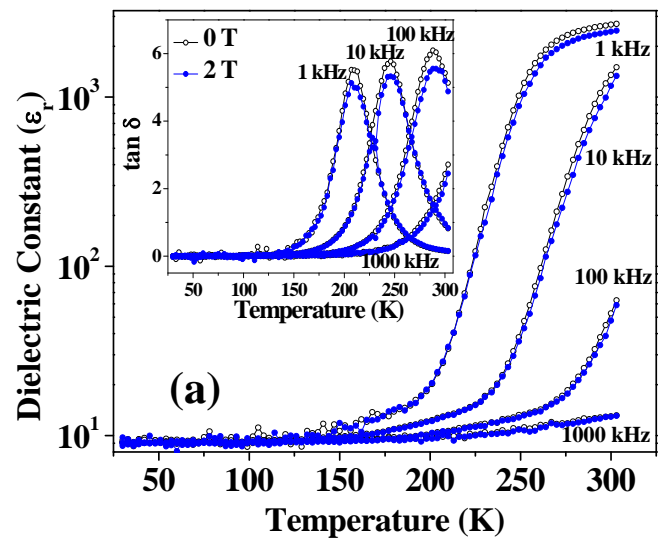

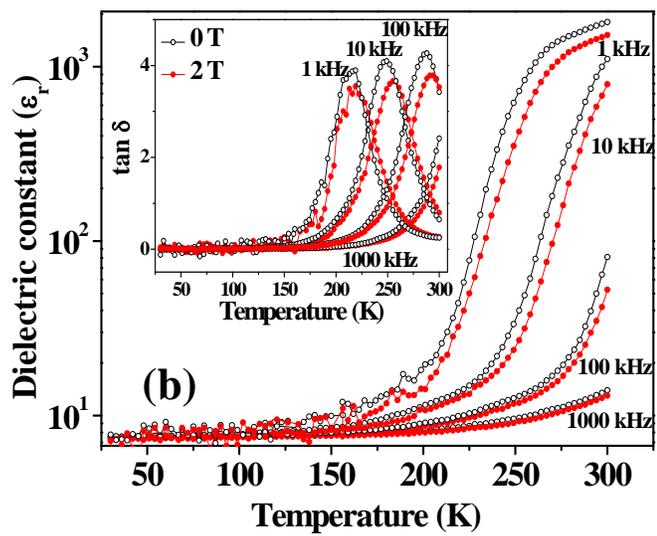

**Figure 2**

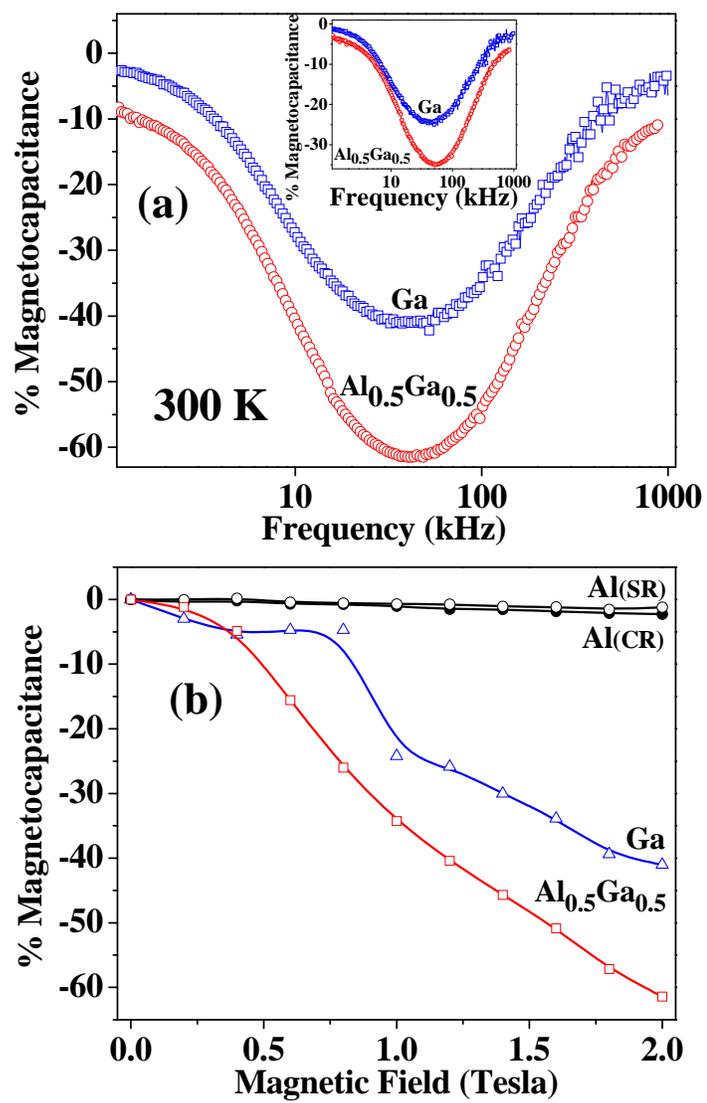

Figure 3